\documentclass[journal]{IEEEtran}

\ifCLASSINFOpdf
  \usepackage[pdftex]{graphicx}
\usepackage{amsmath}
\usepackage{amsthm,amssymb}
\usepackage{booktabs}
\usepackage[table]{xcolor}

\ifCLASSOPTIONcompsoc
 \usepackage[caption=false,font=normalsize,labelfont=sf,textfont=sf]{subfig}
\else
 \usepackage[caption=false,font=footnotesize]{subfig}
\fi
\usepackage{algorithmic}
\usepackage{gensymb}
\usepackage[]{algorithm2e}

\usepackage{tabularx}
\usepackage{blindtext}
\usepackage{cite}
\usepackage{cleveref}
\usepackage{booktabs}
\usepackage{pifont}
\usepackage{url}
\usepackage{etoolbox}
\usepackage[compact]{titlesec}

\newtoggle{long}
\toggletrue{long}
\newcommand{\iflong}[1]{\iftoggle{long}{#1}{}}
\newcommand{\ifelselong}[2]{\iftoggle{long}{#1}{#2}}

\newtoggle{red_flag}
\togglefalse{red_flag}

\usepackage{ifthen}
\newboolean{showcomments}
\setboolean{showcomments}{true}
\newcommand{\zach}[1]{\ifthenelse{\boolean{showcomments}}
{ \textcolor{blue}{(ZL:  #1)}}{}}
\newcommand{\slow}[1]{\ifthenelse{\boolean{showcomments}}
{ \textcolor{red}{(SL:  #1)}}{}}

\newcommand{\beq}{\begin{equation}}
\newcommand{\eeq}{\end{equation}}
\newcommand{\bq}{\begin{eqnarray}}
\newcommand{\eq}{\end{eqnarray}}
\newcommand{\bqn}{\begin{eqnarray*}}
\newcommand{\eqn}{\end{eqnarray*}}
\newcommand{\bee}{\begin{enumerate}}
\newcommand{\eee}{\end{enumerate}}
\newcommand{\bi}{\begin{itemize}}
\newcommand{\ei}{\end{itemize}}

\newcommand{\cmark}{\ding{51}}%
\newcommand{\xmark}{\ding{55}}%

\begin{document}

\title{Adaptive Charging Networks:\\ A Framework for Smart Electric Vehicle Charging}

\author{
\IEEEauthorblockN{
Zachary~J.~Lee,~\IEEEmembership{Graduate~Student~Member,~IEEE,},
George~Lee,
Ted~Lee,
Cheng~Jin,
Rand~Lee,
Zhi~Low,
Daniel~Chang,
Christine~Ortega,
Steven~H.~Low~\IEEEmembership{Fellow,~IEEE}}

\iflong{%
\thanks{
This work was supported in part by the National Science Foundation under the Graduate Research Fellowship Program (Grant No. 1745301), NSF AIR-TT (Grant No. 1602119), NSF ECCS (Grant No. 1932611), and NSF CCF (Grant No. 1637598). It also received support under the Resnick Sustainability Institute Graduate Fellowship, Caltech RocketFund, Caltech CI2 Grant, the Emerging Technologies Coordinating Council of Utilities, and Well Fargo/NREL IN2.}}%
\thanks{Z. Lee, C. Ortega, and S. Low are with the Division of Engineering and Applied Sciences, California Institute of Technology, Pasadena, CA 91125 USA (e-mail: zlee@caltech.edu; slow@caltech.edu).}%
\thanks{R. Lee, D. Chang, and Z. Low were formerly with the California Institute of Technology, Pasadena, CA 91125 USA}%
\thanks{G. Lee, T. Lee, and C. Jin are with PowerFlex, Los Altos, CA, 94022 USA.}%
}


\maketitle

\begin{abstract}
We describe the architecture and algorithms of the Adaptive Charging Network (ACN), which was first deployed on the Caltech campus in early 2016 and is currently operating at over 100 other sites in the United States. The architecture enables real-time monitoring and control and supports electric vehicle (EV) charging at scale. The ACN adopts a flexible Adaptive Scheduling Algorithm based on convex optimization and model predictive control and allows for significant over-subscription of electrical infrastructure. We describe some of the practical challenges in real-world charging systems, including unbalanced three-phase infrastructure, non-ideal battery charging behavior, and quantized control signals. We demonstrate how the Adaptive Scheduling Algorithm handles these challenges, and compare its performance against baseline algorithms from the deadline scheduling literature using real workloads recorded from the Caltech ACN and accurate system models. We find that in these realistic settings, our scheduling algorithm can improve operator profit by 3.4 times over uncontrolled charging and consistently outperforms baseline algorithms when delivering energy in highly congested systems. 
\end{abstract}

\IEEEpeerreviewmaketitle

\section{Introduction}
\IEEEPARstart{T}{he} number of electric vehicles on the road is expected to reach 125 million by 2030, generating 404 TWh of additional electricity demand \cite{bunsen2018global}.
Charging these EVs cleanly, affordably, and without excessive stress on the grid will require advances in charging system design, hardware, monitoring, and control. Collectively we refer to these advances as smart charging. Smart charging will substantially reduce the environmental footprint of transportation while unlocking immense potential for demand-side management.

Smart charging is especially crucial for large-scale charging facilities such as those in workplaces, apartment complexes, shopping centers, airports, and fleet charging facilities. Providing charging at these diverse sites is vital to the widespread adoption of electric vehicles. Doing so can reduce range anxiety and provide an alternative to personal charging ports for those who cannot install them at their homes. Since many of these sites will provide charging during daytime hours, they can make use of abundant solar energy production and enable EVs to provide grid services throughout the day. However, with current technology, most sites are unable to install more than a few charging ports due to limited infrastructure capacity and fear of high electricity bills. Smart charging allows sites to scale their port capacity without costly infrastructure upgrades. Moreover, scheduling algorithms can reduce operating costs by optimizing for time-of-use tariffs, demand charges, and on-site renewable generation.
\iflong{Algorithms can also enable additional revenue streams by providing grid services.}

In this paper, we report the design, implementation, and application of a smart charging system that we call an Adaptive Charging Network (ACN).  In Section~\ref{sec:overview}, we describe the architecture through the lens of the first ACN, which was built on the Caltech campus in 2016, shown in Fig.~\ref{fig:ACN_photo}. The ACN enables real-time control and monitoring of charging systems at scale. It has spawned a company, PowerFlex Systems, which operates over 100 similar charging systems around the United States. \iflong{These include national laboratories, universities, schools, businesses, apartment complexes, hotels, and public parking facilities.}

Through our experience building large-scale charging facilities, we find that common assumptions made in theoretical models do not hold in many practical systems. We describe these in Section~\ref{sec:key-insights}. This makes it challenging to apply algorithms proposed in the literature directly. In order to develop practical and robust algorithms for the ACN, we present the Adaptive Scheduling Algorithm (ASA) framework for online (causal) smart charging algorithms based on convex optimization and model predictive control (MPC). We describe the ASA framework in Section~\ref{sec:scheduling-framework}. Then, in Section~\ref{sec:applications}, we demonstrate ASA's performance in the context of maximizing energy delivery in congested infrastructure and minimizing operating costs, including demand charge using simulations based on real data collected from the ACN.

\begin{figure}[!t]
\includegraphics[width = \columnwidth]{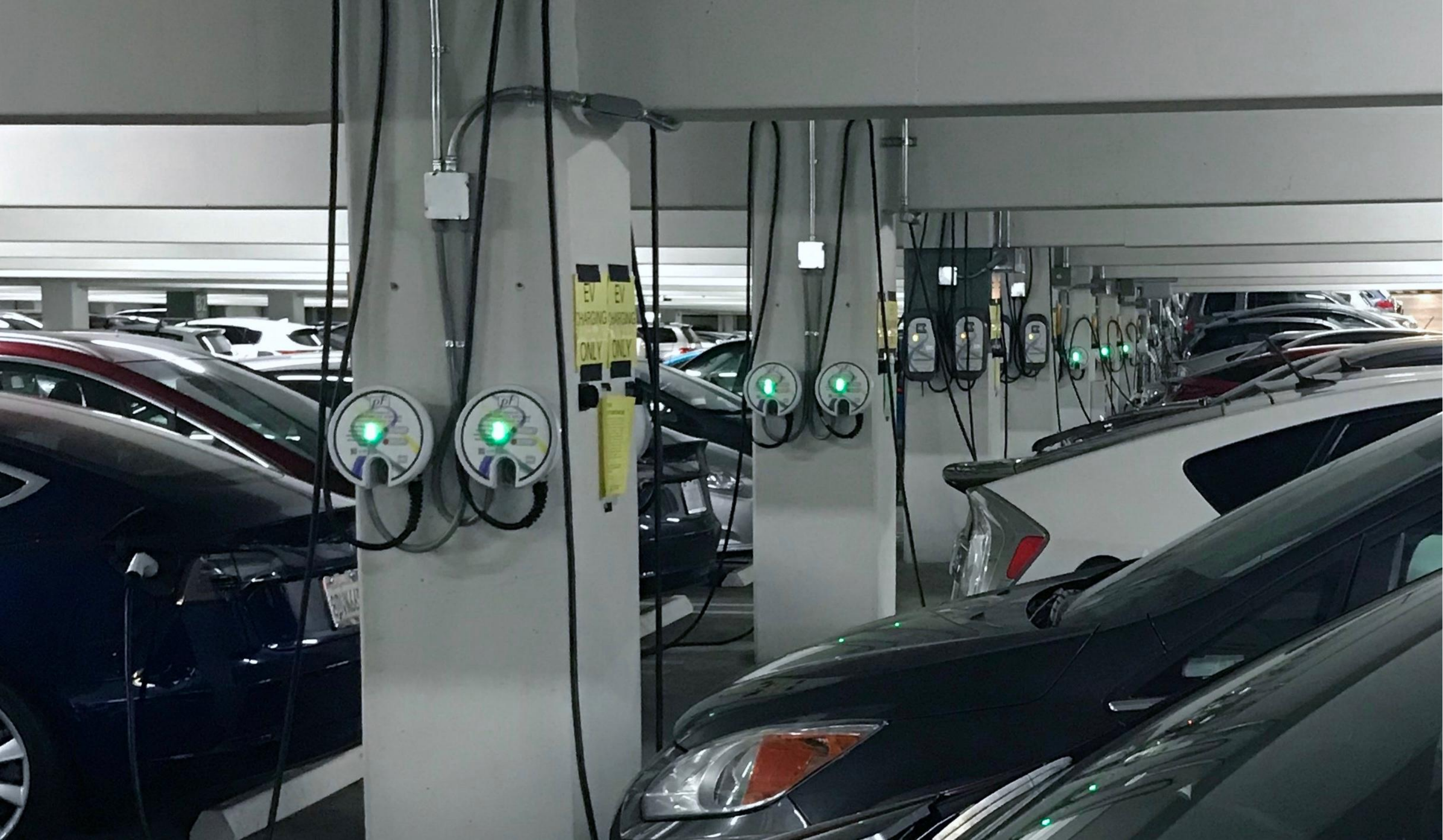}
\centering
\caption{The ACN smart EV charging testbed at Caltech.}
\label{fig:ACN_photo}
\end{figure}

Beyond serving as a model for smart charging systems, the ACN has led to the creation of the ACN Research Portal. This portal has three parts: ACN-Data, a collection of real fine-grained charging data collected from the Caltech ACN and similar sites \cite{lee_ACN-Data_2019}; ACN-Sim, an open-source simulator which uses data from ACN-Data and realistic models derived from actual ACNs to provide researchers with an environment to evaluate their algorithms and test assumptions \cite{lee_acnsim_2019}; and ACN-Live, a framework for safely field testing algorithms directly on the Caltech ACN. Thus the ACN has proven to be a valuable tool in both commercial and academic environments. 

\section{Related Work}
Several smart EV charging systems have been developed, though usually at a smaller scale than the ACN. The Smart Energy Plaza (SEP) at Argonne National Laboratory consists of six controllable level-2 EVSEs\cite{bohn_real_2016}. Likewise,  the WinSmartEV system at UCLA consists of quad-port EVSEs capable of sharing a single oversubscribed circuit and multiple level-1 chargers with binary control \cite{chynoweth_smart_2014, chung_master-slave_2014}. My Electric Avenue tested a control system to managed charging from 200 Nissan LEAFs to manage congestion in the distribution system \cite{quiros-tortos_how_2018}. The Parker project utilized a testbed of 10 bi-directional EVSEs at a commercial site to investigate the potential of EVs to provide frequency regulation services and adapt to marginal emissions signals \cite{andersen_parker_2019}.

There is also a tremendous literature on EV charging algorithms; see surveys \cite{wang_smart_2016, mukherjee_review_2015} for extensive pointers to the literature. 
Some recent work specifically related to large-scale EV charging includes \cite{chen_iems_2012,yu_intelligent_2016, wang_two-stage_2016, wang_predictive_2017, nakahira_smoothed_2017, zhang_optimal_2017, frendo_real-time_2019, alinia_online_2020}.

Several works have been based on the ACN and data collected from it. \cite{lee_ACN-Data_2019} uses data from the ACN to predict user behavior, size solar generation for EV charging systems, and evaluate the potential of EV charging to smooth the duck curve. \cite{sun_electric_2020} clusters sessions based on their charging behavior using time series of charging current collected from the ACN. \cite{lee_pricing_2020} proposes a pricing scheme to allocate costs (including demand charge) to charging sessions. \cite{li_real-time_2020, schlund_flexability_2020} use data to quantify fleet-level flexibility within a charging facility, and \cite{al_zishan_adaptive_2020} uses ACN-Sim to train reinforcement learning agents to schedule large-scale EV charging. 

\iftoggle{long}{Our work adds to the existing body of research insights from building real-world EV charging systems. It}
{This work}
extends \cite{lee_adaptive_2016, lee_large-scale_2018}, which describe the ACN in earlier stages of development. It provides a more thorough description of the ACN
\iflong{architecture and algorithms}{and ASA} 
and demonstrates ASA's effectiveness in realistic scenarios through simulation.
\section{Adaptive Charging Network Architecture}\label{sec:overview}
We first describe the architecture of the ACN through the lens of the Caltech ACN. Since its installation in early 2016, the Caltech ACN has grown from 54 custom-built level-2 electric vehicle supply equipment (EVSEs) in a single garage \cite{lee_adaptive_2016}, to 126 commercially available level-2 EVSEs, one 50 kW DC Fast Charger (DCFC), and four 25 kW DCFC, spread between three parking garages on campus. These EVSEs have delivered over 1,103 MWh as of July 7, 2020. The Caltech ACN is a cyber-physical system, as shown in Fig.~\ref{fig:ACN_arch}, that consists of five interacting subsystems: (1) the information system which is responsible for collecting information and computing control actions; (2) the sensor system which gathers information from the physical system; (3) the actuation system (made up of EVSEs and the EVs' battery management systems) which controls each vehicle's charging rate; (4) the physical system (electrical infrastructure) which delivers power to the EVs and other loads within the system; (5) drivers who provide data to the system and decide when their vehicles are available to charge. 
\ifelselong{In this section, we describe these subsystems and their interactions.}{We now focus on the components of the information and physical systems in detail.}

\begin{figure}[!t]
\includegraphics[width =\columnwidth]{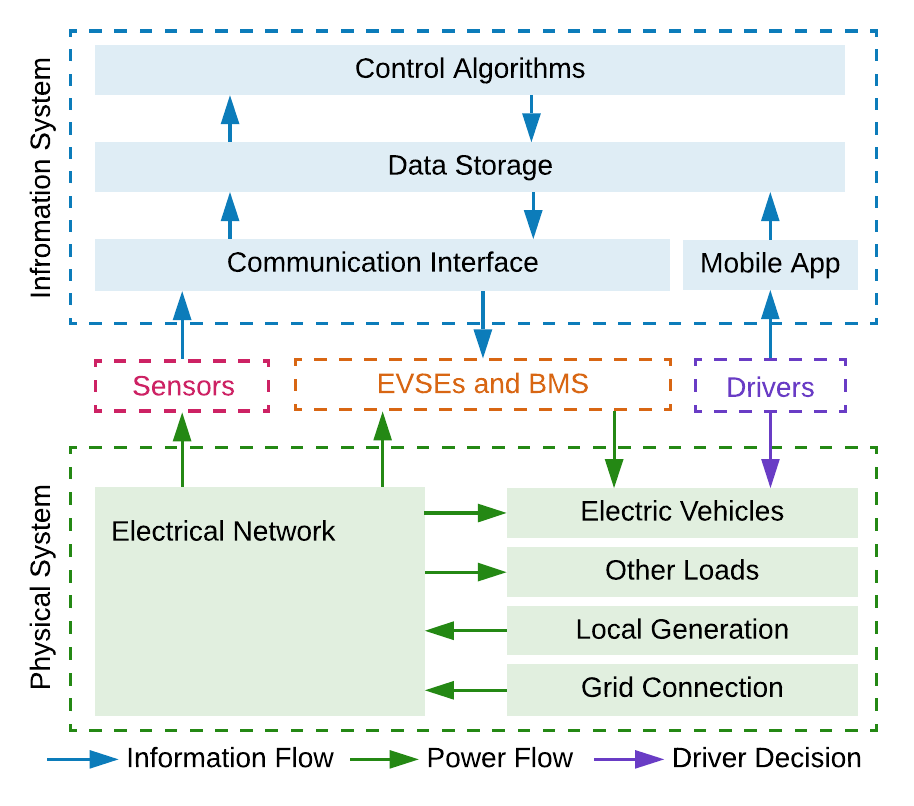}
\centering
\vspace{-0.35in}
\caption{Architecture of the ACN. Blue and green arrows signify the flow of information and power, respectively. Sensors measure power flowing in the electrical network and convert this into information. Likewise, EVSEs and the EVs' onboard battery management system (BMS) work together as actuators to control the flow of power into each EVs battery based on signals from the information system. Drivers provide information to the system via a mobile app and directly control when EVs are plugged in or unplugged from the system (signified by the purple arrow).}
\label{fig:ACN_arch}
\end{figure}
\subsection{Information system}
ACN's information system collects and stores relevant data and computes control actions.
It consists of four components:

\textbf{Communication interface:}
The communication interface collects sensor data and passes it to the data storage layer. It also passes signals generated by the control algorithms to the corresponding EVSEs. An industrial computer within the parking garage controls this communication interface. It connects to the cloud-based components through a cellular internet connection and to sensors and EVSEs within the ACN via a Zigbee based mesh network.

\textbf{Data storage:}
The ACN utilizes a relational database to store information such as site configurations, driver profiles, and charging session parameters. A dedicated time-series database stores measurements like voltage, current, and power readings taken from sensors in the electrical network
%
. 
The data storage layer allows us to create visualizations for drivers and site operators%
\iflong{, which helps them understand the state of the system and their own EV's charging trajectory in real-time}, as seen at \cite{caltech_dashboard_2019}.

\textbf{Mobile app:} 
Our mobile app collects data directly from drivers. After setting up an account, a driver scans a QR code on the EVSE, and then provides an estimated departure time and requested energy. If the driver's plans change, they can update these parameters throughout the charging session. The app also allows the site to collect payment and, if desired, implement access control. To ensure that drivers provide information through the app, an EV will only charge at 8 amps until the driver claims the session through the app. After 15 minutes, if the session is not claimed, it will be terminated, and the EVSE will cease charging.

\textbf{Control algorithms:}
The control layer takes inputs from the data layer and calculates a charging schedule for each EV in the system.
We use an event-based system to trigger the scheduling updates. The events considered include a vehicle plugging in or unplugging, a driver changing request parameters, or a demand response signal from the utility. Events are handled by a publish-subscribe model.
Whenever an event occurs, or the time since the last charging schedule update exceeds a threshold (for example, 5 min), we compute a new charging schedule. These periodic computations close the control loop and account for discrepancies between the control signal sent to each EV and its actual current draw. We describe this model predictive control framework in detail in Section~\ref{sec:scheduling-framework}.

\iflong{%
\subsection{EVSEs and Battery Management System}\label{sec:evse}
To control charging rates, we use the pilot signal mechanism defined by the J1772 standard for level-2 EVSEs \cite{sae_sae_2017}. According to this standard, the EVSE can communicate an upper bound to the EV's battery management system (BMS) that limits the amount of current it may draw from the EVSE. Because it is only an upper bound, the vehicle's BMS may choose to charge at a lower rate. This can occur for various reasons such as the pilot signal being higher than the vehicle's maximum charging rate or the BMS limiting current draw as the battery reaches a high state of charge.  It can be difficult to diagnose why a car is charging below its allocated pilot signal since the J1772 standard does not provide a way to gather the EV's state of charge. Also, most EVSEs on the market today, including the ClipperCreek, Webasto, and Tesla EVSEs in the Caltech ACN, only support a finite set of pilot signal values and require quantization of the control signal.
}

\iflong{%
\subsection{Sensors}
Sensors provide a bridge between the physical system and the information system. These sensors measure power, current, and voltage within the local electrical network, allowing us to monitor the system state and accurately track energy usage. The sensors also provide feedback for the control algorithm.
}

\subsection{Physical system} \label{sec:physical_system}
The physical system of the ACN includes the local electrical network (including transformers, lines, breakers, loads, and local generation), a connection to the grid, and the electric vehicles. Fig.~\ref{fig:network-top} shows the topology of the local electrical network for one garage of the Caltech ACN. Power is delivered to the garage from the distribution transformer via three-phase service at 480 V\textsubscript{LL}. 

From there, power is distributed throughout the garage via the main switch panel. The ACN is connected to this panel by two 150 kVA delta-wye transformers $t_1$ and $t_2$, which step the voltage down to 120 V\textsubscript{LN}. Each level-2 EVSE is a single-phase load connected line-to-line (208 V\textsubscript{LL}) with a maximum current draw between 32 A to 80 A depending its type. Because of unequal loading between phases,
\iflong{which is unavoidable due to the stochastic nature of driver demands on the system,}
balanced operation cannot be assumed. This makes protection of transformers $t_1$ and $t_2$ challenging which we discuss in Section~\ref{sec:three_phase_model}. 
\iflong{Another interesting feature of the Caltech ACN is the two pods of eight EVSEs. These pods are each fed by an 80 amp line. Since each EVSE in the pod has a maximum charging rate of 32 A, these lines are oversubscribed by 3.2 times. This demonstrates how smart charging can allow sites to scale EVSE capacity with existing infrastructure.}

\iflong{%
In addition to the 78 EVSEs in the garage, the ACN also includes a 50 kW DC fast charger (DCFC). This DCFC is a balanced three-phase load.
While this garage does not have local generation, other garages in the Caltech ACN and other PowerFlex sites have on-site solar generation.}
\begin{figure}[!t]
\includegraphics[width = \columnwidth]{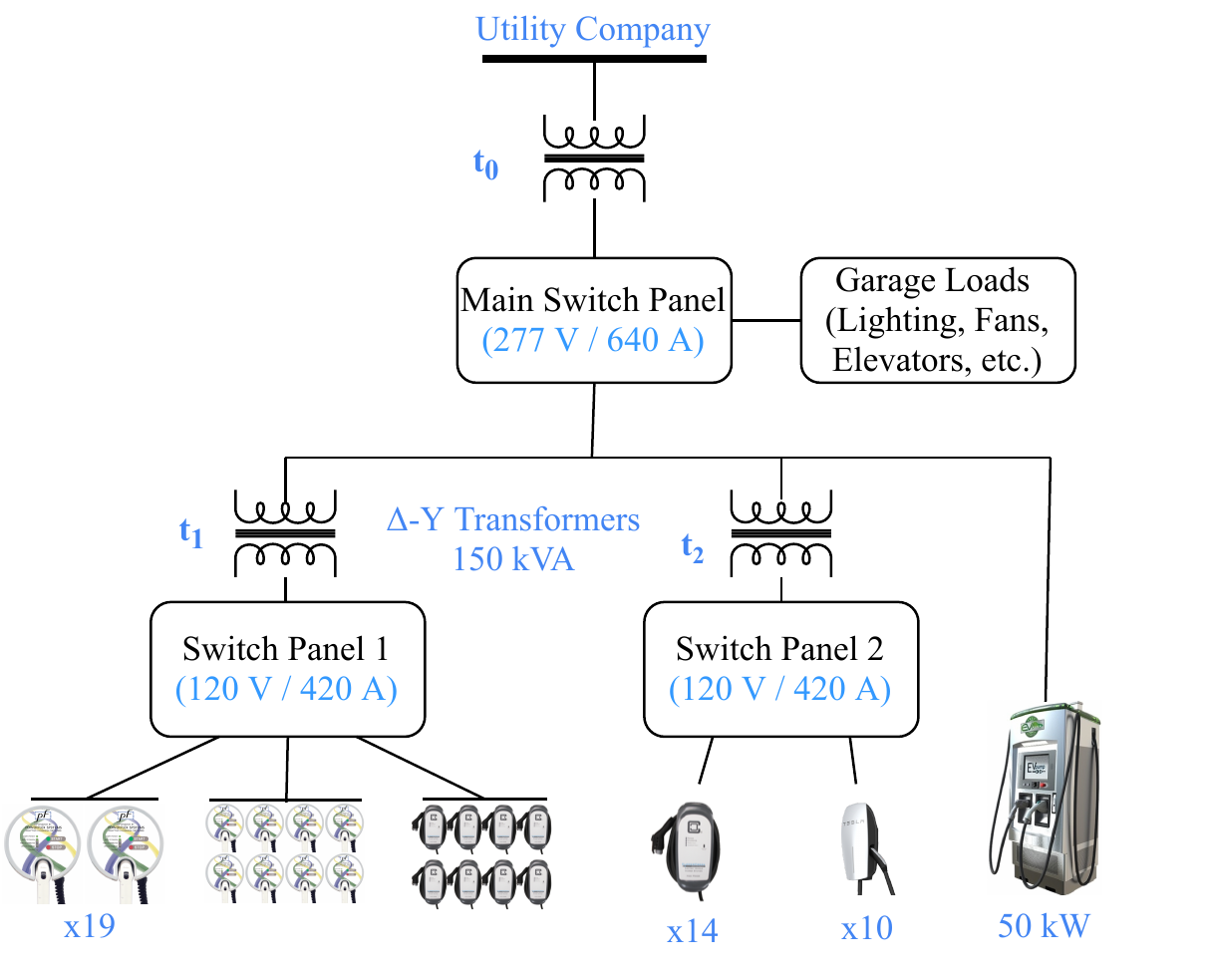}
\centering
\vspace{-0.35in}
\caption{System topology for the California Parking Garage in the Caltech ACN. The system consists of 78 EVSEs and one 50 kW DC Fast Charger. Switch Panel 1 is fed by a 150 kVA transformer and feeds 54 6.7 kW EVSEs, leading to a 2.4X over-subscription ratio. Nineteen of these lines feed pairs of two 6.7 kW AeroVironment EVSEs. Two additional 80 A lines feed pods of eight 6.7 kW EVSEs each, one pod of AeroVironment stations and the other of Clipper Creek stations. Switch Panel 2 is fed by an identical 150 kVA transformer and feeds 14 13.3 kW Clipper Creek EVSEs and 10 16.6 kW Tesla EVSEs. Each of these EVSEs has a dedicated 80 A line. All EVSEs in the system are connected line-to-line at 208 V. The 50kW DC Fast Charger from BTC Power is a balanced 3-phase load connected directly to the main switch panel. We do not directly control the DCFC at this time.
}
\label{fig:network-top}
\end{figure}

\iflong{
\subsection{Drivers}
Human behavior can add significant randomness to the system. Drivers may arrive, depart, or change their input parameters at any time. Drivers are also difficult to model. Input through the mobile app can be highly inaccurate, as shown in \cite{lee_ACN-Data_2019}. To combat this, we have explored using machine learning to predict driver parameters \cite{lee_ACN-Data_2019} as well as pricing schemes that incentivize drivers to provide accurate estimates \cite{lee_pricing_2020}.}


\section{Practical Challenges from the Testbed} \label{sec:key-insights}
By building and operating the Caltech ACN, we have identified several important features of the physical system that have not been addressed in the EV charging literature but pose real problems for implementing practical EV scheduling algorithms. 
Among these are proper modeling of the unbalanced three-phase electrical network, incorporation of EVSE quantization, and adaptation to non-ideal battery behavior.
We describe these models in this section and explain how we incorporate them into an MPC framework in Section~\ref{sec:scheduling-framework}. These models also form the basis of the component models included in ACN-Sim \cite{lee_acnsim_2019}.

\subsection{Infrastructure modeling}\label{sec:three_phase_model}
As discussed in Section~\ref{sec:physical_system}, the electrical infrastructure within the ACN is oversubscribed and often unbalanced. In our data, we observe that without proper control, these phase imbalances can be significant%
\iflong{, as seen in Fig.~\ref{fig:phase_imbalance}}. 
While many algorithms have been proposed to handle charging with an aggregate power limit or even a hierarchy of limits, most previous work has focused on single-phase or balanced three-phase systems, making them inapplicable to charging systems like the ACN. An exception to this is the work of De Hoog et al. \cite{de_hoog_optimal_2015}, which considers an unbalanced three-phase distribution system but only in the case of wye-connected EVSEs. In contrast, the EVSEs in the ACN and most large charging systems in the United States are connected line-to-line%
\iflong{, as shown in Fig.~\ref{fig:circuit-diag}}.  

\iflong{%
\begin{figure*}[htbp]
\includegraphics[width =0.88\textwidth]{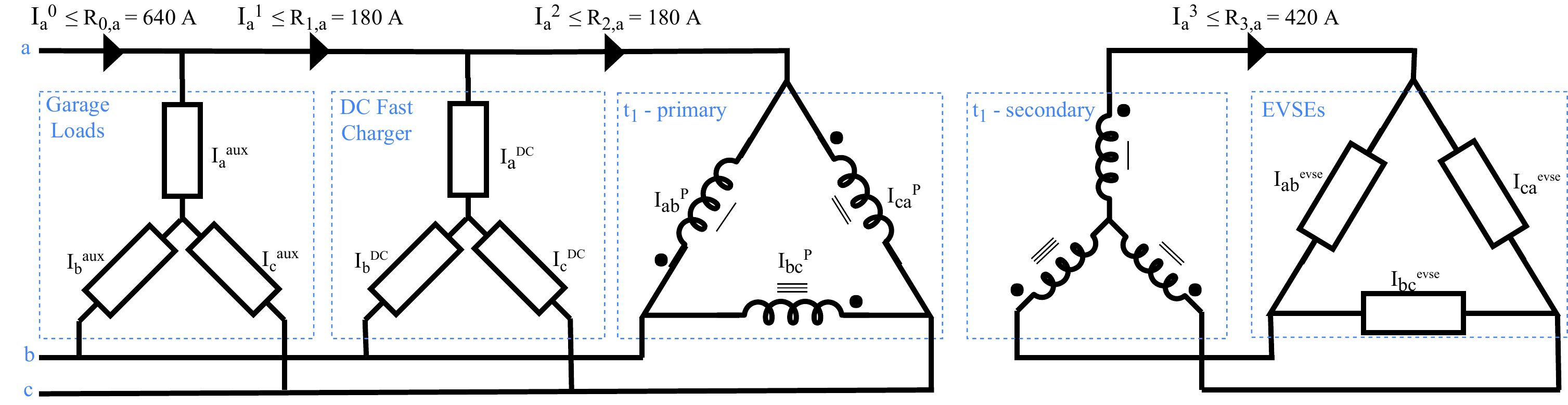}
\centering
\caption{Circuit diagram depicting the connection loads within the California Parking Garage. For simplicity, transformer $t_2$ is omitted and all EVSEs between phases A and B have been lumped together as $I_{ab}^{evse}$, and so forth for BC and CA.}
\label{fig:circuit-diag}
\end{figure*}
}

\iftoggle{long}{%
 \begin{figure}[!t]
    \centering
    \includegraphics[width=\columnwidth]{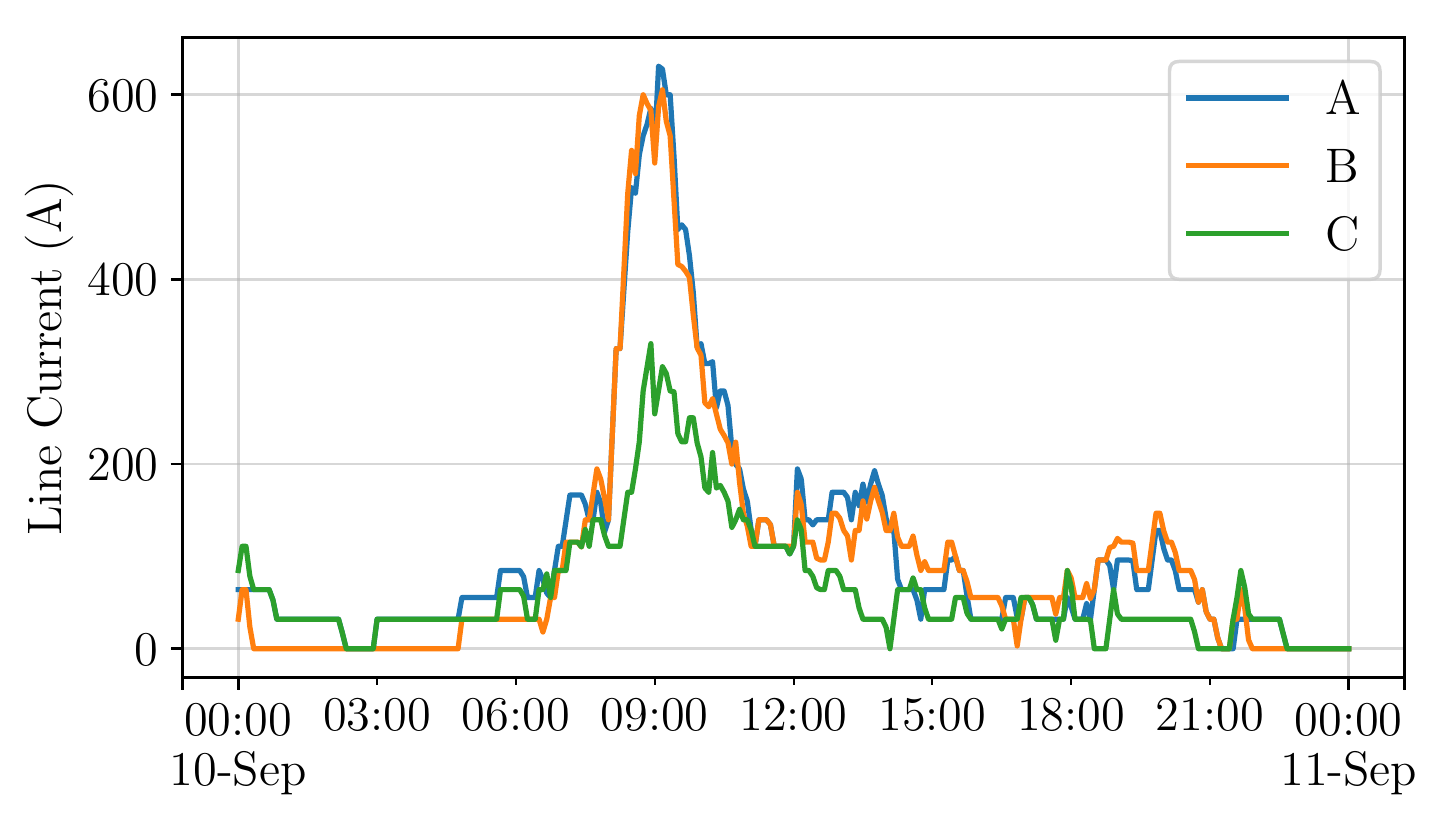}
    \caption{Line currents from uncontrolled charging. We note significant current imbalances caused by differences in the allocation of EVSEs to phases and driver preferences. In the ACN, phase AB has 26 stations, whereas phases BC and CA each have 14. This imbalance is caused by the two 8-EVSE pods which are both on phase AB.}
    \label{fig:phase_imbalance}
\end{figure}
}

In general, infrastructure constraints can be expressed as upper bounds on the magnitudes of currents within the system. By Kirchoff's Current Law, we can express any current within the system as the sum of load currents. Let $\mathcal{V}(t)$ denote the set of EVs which are available to be charged at time $t$. The current draw of EV $i$ at time $t$ can be expressed as a phasor in the form $r_i(t) e^{j\phi_i}$ where $r_i(t)$ is the charging rate of the EV in amps, and $\phi_i$ is the phase angle of the current sinusoid. We assume in this model that each charging EV has a unity power factor, so $\phi_i$ is known based on how the EVSE is connected and the voltage phase angles (which we assume are separated by $\pm 120\degree$).\footnote{This is reasonable since EVs' onboard charger generally includes power factor correction and voltage phase angles can be easily measured.} We can then model constraints within the electrical system as
\begin{equation}\label{eq:three_phase_model}
    \left| \sum_{i \in \mathcal{V}} A_{li} r_i(t)e^{j\phi_i} + L_l(t)\right| \leq \ c_{lt} \quad \quad t \in \mathcal{T}, l \in \mathcal{L}
\end{equation}
\noindent Infrastructure limits of the network are indexed by resources $l\in \mathcal{L}$, e.g., $l$ may refer to a transformer or a breaker on a phase.
For each constraint $l$, $c_{lt}$ is a given capacity limit for time $t$ and $L_l(t)$ is the aggregate current draw from uncontrollable loads through resource $l$. $A=(A_{li}) \in \mathcal{R}^{|\mathcal{L}|\times|\mathcal{V}|}$ is a matrix which maps individual EVSE currents to aggregate currents within the network. Matrix $A$ can account for both the connection of loads and lines as well as the effect of transformers, such as the delta-wye transformers in the Caltech ACN. The constraints in (\ref{eq:three_phase_model}) are second-order cone constraints, which are convex and can be handled by many off-the-shelf solvers such as ECOS, MOSEK, and Gurobi. In some applications, however, these constraints could be too computationally expensive or difficult to analyze. Simpler, but more conservative constraints can be derived by observing
\begin{equation*}
    \left| \sum_{i \in \mathcal{V}} A_{li} r_i(t) e^{j\phi_i} \right| \ \leq \sum_{i\in\mathcal{V}} |A_{li}| r_i(t)
\end{equation*}
This yields conservative affine constraints in the form
\begin{equation}\label{eq:inf_constraints_affine}
    \sum_{i \in \mathcal{V}} |A_{li}| r_i(t) + |L_l(t)| \ \leq \ c_{lt} \quad t \in \mathcal{T}, l \in \mathcal{L}
\end{equation}

For an example on how to find $A$ specifically for a subset of the Caltech ACN network, as well as a comparison of the performance of \eqref{eq:three_phase_model} and \eqref{eq:inf_constraints_affine}, see \cite{lee_large-scale_2018}.

\iflong{%
\subsection{Battery Management System behavior} \label{sec:bms_model}
For level-2 EVSEs, each EV's onboard charger and battery management system (BMS) controls its charging rate. As discussed in Section~\ref{sec:evse}, the EV's actual charging current often deviates, sometimes significantly, from the pilot signal it receives. This requires us to develop algorithms that accurately model battery behavior or are robust against deviations from simpler models. While many tractable models for battery charging behavior exist, these models require information about the specific battery pack and initial state of charge of the vehicle \cite{kazhamiaka_simple_2018, kazhamiaka_tractable_2019}. Other models rely on machine learning to learn the relationship between state of charge and current draw \cite{frendo_data-driven_2020}.
However, these ML models still require access to the state of charge of the vehicle. Since this information is not available with current charging hardware, we use a model-free approach to estimate battery behavior in real-time and use closed-loop control to account for modeling errors. This approach is described in Section~\ref{sec:battery-tail-reclamation}.
}

\subsection{EVSE limitations}\label{sec:evse_limits}
In practice, EVSEs impose limits on the pilot signals which they support. For example, the J1772 standard does not allow pilot signals below 6 A (except 0). Also, most commercially available EVSEs only support a discrete set of pilot signals. Within the Caltech ACN, we have four types of EVSEs. EVSEs from ClipperCreek only support five pilot signals \{0, 8, 16, 24, 32\} for 32 amp EVSEs and \{0, 16, 32, 48, 64\} for 64 amp EVSEs. EVSEs from Webasto, Tesla, and OpenEVSE offer more control with 1 A (Webasto) or 0.1 A (Tesla and OpenEVSE) increments between 6 A and their maximum rate (32 A for Webasto and 80 A for Tesla and OpenEVSE). These limitations can be expressed mathematically as:
\begin{equation*}  
r_i(t) \in \rho_i(t) \quad \forall i,t
\end{equation*}
where $\rho_i(t)$ denote the set of allowable charging rates for EV $i$ at time $t$, which can depend on both the EVSE and our model of the EV's BMS%
\iflong{ described in Section~\ref{sec:bms_model}}. 

We also require the charging rate to be non-zero from when a car plugs in until its charging demand is met. 
This constraint helps prevent contactor wear in the EVSEs and improves user experience, since most vehicles will notify their owner when charging stops.
We can encode this constraint as:
\begin{equation} \label{eq:flapping-rate-const}
  r_i(t) \in
  \begin{cases}
    \rho_i(t) \setminus \{0\} & \text{if $\sum_{t=1}^{T} r_i(t) < \, e_i$}\\
    \{0\} & \text{otherwise}
  \end{cases}
\end{equation}
where $e_i$ is the energy request of EV $i$.
Unfortunately, these constraints are discrete, making it difficult to incorporate them into optimization-based algorithms. In section \ref{sec:post_processing} we propose heuristics to deal with these discrete constraints.

\section{Online Scheduling Framework} \label{sec:scheduling-framework}
To address these challenges, we have developed a practical and flexible framework for online scheduling of EV charging based on model predictive control and convex optimization. Within this
framework, we introduce constraints that address unbalanced three-phase infrastructure and utilize feedback to account for inaccuracies in modeling, such as non-ideal battery behavior. Finally, we introduce a heuristic approach to account for discrete constraints arising from EVSE limitations efficiently.

\subsection{Model predictive control}\label{sec:MPC}
The ACN computes charging rates using model predictive control, described in Alg. \ref{alg:MPC}.
\begin{algorithm}[htbp] \label{alg:MPC}
\SetAlgoLined
\SetNlSty{texttt}{(}{)}
\For{$k \in \mathcal{K}$}{
    \nl $\mathcal{V}_k := \{i\in \hat{\mathcal{V}}_k \mid e_i(k) > 0$ \textbf{AND} $d_i(k) > 0$\} \label{alg:active_set}\\
     \nl \If{event fired \textbf{OR} time since last computation $> P$ \label{alg:recomp_condition}}
     {
     \nl $(r^*_i(1),...,r^*_i(T), i \in \mathcal{V}_k)$ := \textbf{OPT}$(\mathcal{V}_k, U_k, \mathcal{R}_k)$ \label{alg:scheduler.1} \\
    \nl $r_i(k+t) := r^*_i(1+t), \ t=0, \dots, T-1$ \label{alg:scheduler.2}
    }
    \nl set the pilot signal of EV $i$ to $r_i(k)$, $\forall i \in \mathcal{V}_k$ \label{alg:pilot_signal}\\
   \nl $e_i(k+1) := e_i(k) - \hat{e}_i(k)$, $\forall i \in \mathcal{V}_k$ \label{alg:energy_update}\\         \nl $d_i(k+1) := d_i(k) - 1$, $\forall i \in \mathcal{V}_k$ \label{alg:departure_update}
 }
 \caption{Adaptive Scheduling Algorithm (ASA)}
\end{algorithm}
We use a discrete time model, with time indexed by $k$ in $\mathcal{K} := \{1,2,3,...\}$. The length of each time period is $\delta$ e.g. 5 minutes. At time $k$, $\hat{\mathcal{V}}_k$ is the set of all EVs present at the ACN and $\mathcal{V}_k \subseteq \hat{\mathcal{V}}_k$ is the subset of \emph{active} EVs, i.e. the set of EVs whose energy demands have not been met.
The state of EV $i\in\mathcal V_k$ at time $k$ is described by a tuple ($e_i(k)$, $d_i(k)$, $\bar{r}(k)$) where $e_i(k)$ is the remaining energy demand of the EV at the beginning of the period, $d_i(k)$ is the remaining duration of the session, and $\bar{r}(k)$ is the maximum charging rates for EV $i$. In addition, we define $\hat{e}(k)$ to be the measured energy delivered to the EV over time interval $k$. For simplicity of notation, we express $r_i(t)$ in amps and $e_i(t)$ and $\hat{e}_i(t)$ in $\delta$ $\times$ amps assuming nominal voltage.

We now describe the MPC algorithm. In line \ref{alg:active_set} we compute the active EV set
$\mathcal{V}_k$ by looking for all EVs currently plugged in which have non-zero remaining energy demand and are not already scheduled to depart. 
We then check, in line \ref{alg:recomp_condition}, if we should compute a new optimal schedule. This is done whenever an event-fired flag is True, or when the time since the last computed schedule exceeds $P$ periods.

If a new schedule is required, we call the optimal scheduling algorithm $\textbf{OPT}(\mathcal{V}_k, U_k, \mathcal{R}_k)$ in line \ref{alg:scheduler.1} that takes the form:
\begin{subequations}
\bq
\max_{\hat r} & & U_k(\hat r)
\\
\text{s.t.} & & \hat r \in \mathcal R_k
\eq
\label{eq:SCH.1}
\end{subequations}
The set $\mathcal V_k$ of active EVs defines the optimization variable $\hat r:=(\hat r_i(1), \dots, \hat r_i(T), i\in\mathcal V_k)$ for every active EV $i$ over the optimization horizon $\mathcal T := \{1, \dots, T\}$.
The utility function $U_k$ encodes the problem's objective while the feasible set $\mathcal{R}_k$ encodes various constraints. They will be discussed in detail in the next two subsections.
Note that \textbf{OPT} does not have a notion of the current time $k$ and returns an optimal solution $r^*_i := (r_i^*(1),...,r_i^*(T))$ of \eqref{eq:SCH.1} as a $T$-dimensional vector for each active EV $i$.
The algorithm then adjusts the indexing and sets the scheduled charging rates of EVs $i$ at time $k$ as $r_i(k+t) := r^*_i(1+t)$, $t = 0,...,T-1$ in line \ref{alg:scheduler.2}.
At every time $k$, regardless of if a new schedule was produced, we set the pilot signal of each EV $i$ to $r_i(k)$ (line \ref{alg:pilot_signal}) and update the system state (lines \ref{alg:energy_update}, \ref{alg:departure_update}) for the next time period.

We now describe how to design the utility function $U_k$ to achieve desirable features and how to model various constraints that define the feasible set $\mathcal R_k$ for practical systems.

\subsection{Utility Functions $U_k$} \label{sec:cost_funct}
In general, charging system operators may have many objectives they wish to achieve via smart charging, including charging vehicles as quickly as possible, maximizing their operating profit, \ifelselong{}{or} utilizing renewable energy sources\iflong{, or smoothing their total load profile}. Operators also have secondary objectives such as fairly distributing available capacity.

To allow operators to specify multiple objectives, our utility function $U_k(r)$ is a weighted sum of utility functions $u_k^v(r)$:
\begin{equation*} 
U_k(r) :=  \sum_{v=1}^V \alpha^{v}_k u_{k}^{v}(r)
\end{equation*}
We allow the utility function to change for each computation.
 Here $u_k^{v}(r)$,~$v=1,...,V$ are a set of utility functions which capture the system operator's objectives and promote desirable properties in the final schedule.
Meanwhile, $\alpha_k^{v} > 0$,~$v=1,...,V$ are time-dependent weights used to determine the relative priority of the various components.
To simplify notations, we will henceforth drop the subscript $k$ when we discuss the computation at time $k$.

\textbf{Charging quickly:}
One common operator objective is to charge all vehicles as quickly as possible. This can be done by specifying an objective such as
\begin{equation*} 
u^{QC}(r) := \sum_{t \in \mathcal{T}} \frac{T-t+1}{T} \sum_{i \in \mathcal{V}} r_i(t)
\end{equation*}
where the reward for delivering energy is strictly decreasing in time.

\textbf{Minimizing cost / maximizing profit:}
Another common objective for system operators is to maximize their operating profit. Let $\pi$ be the per unit revenue from charging and $c(t)$ be the time-varying cost of one unit of energy. 
To account for other loads and generation which share a meter with the ACN, we define the net load 
\begin{equation*}
    N(t) := \sum_{i \in \mathcal{V}} r_i(t) + L(t) - G(t)
\end{equation*}

\noindent where $L(t)$ denotes the net draw of the other loads while $G(t)$ denotes on-site generation such as PV. Since $L(t)$ and $G(t)$ are unknown for $t > 0$ this formulation relies on a prediction of these functions into the future. There are several methods for load/generation prediction proposed in the literature, but these are outside the scope of this paper. We can express the objective of maximizing profit as:

\begin{equation*}\label{eq:profit_max_util_energy_only}
u^{EC}(r) := \pi \sum_{\substack{t \in \mathcal{T}\\i \in \mathcal{V}}} r_i(t) - \sum_{t \in \mathcal{T}} c(t)N(t)
\end{equation*}
This is equivalent to cost minimization when $\pi = 0$. 

\textbf{Minimizing demand charge:}
In addition to energy costs, utilities often impose a price on the maximum power draw in a billing period called demand charge. Since, demand charge is assessed over an entire month, while the optimization horizon is typically $<12$ hours, we replace the full demand charge $P$ with a proxy $\hat{P} \leq P$. We also introduce $q_0$ to be the highest peak so far in the billing period, and $q'$ as a prediction of the optimal peak. The demand charge can then be expressed as:

\begin{equation*} \label{eq:profit_max_util}
        u^{DC}(r) \: := -\hat P\cdot \max \left( \max_{t\in\mathcal{T}}  N(t),\, q_0, q'\right)
\end{equation*}
Note that $\hat{P}$ and $q'$ are tunable parameters. We describe the selection of these in Section~\ref{sec:cost_min}.

\iflong{%
\textbf{Minimizing total load variations:}
Another common objective for EV charging operators is to minimize load variations. We can express this objective as:

\begin{equation*}
        u^{LV}(r) \: := -\sum_{t \in \mathcal{T}} N(t)^2 
\end{equation*}
}

\textbf{Fairly distributing capacity:}
The utility functions described so far are not strictly concave in $r$ and hence the optimal solution, $r^*$, is generally non-unique. We can force a unique optimal solution by including the regularizer:
\begin{equation*}\label{eq:sharing}
u^{ES}(r) := -\sum_{\substack{t \in \mathcal{T}\\ i \in \mathcal{V}}} r_i(t)^2
\end{equation*}
This regularizer also promotes equal sharing among the EVs, which is desirable for the operator and drivers and minimizes line losses along the lines which feed each EVSE. This property comes from the fact that all things being equal, this component is maximized when all charging rates are as low as possible.  Thus, it is sub-optimal to have one EV charging faster than another if both charging at an equal rate would result in the same optimal value for all other objective components. 

\iflong{%
\textbf{Non-completion penalty:}\label{sec:non-completion}
A general goal of EV charging systems is to meet users' energy needs by their deadlines. While this can be accomplished by an equality constraint in $\mathcal{R}_k$, doing so can lead to infeasibility. 
Instead, we can use the inequality constraint \eqref{eq:Constraints.1c}, and add a non-completion penalty of the form:

\begin{equation*}\label{eq:non-completion}
u^{NC}(r) := - \sqrt[p]{\sum_{i \in \mathcal{V}} \left|\sum_{t \in \mathcal{T}} r_i(t) - e_i \right|^p}
\end{equation*}

\noindent where $p \geq 1$.  This is the p-norm of the difference between the energy delivered to each EV and its requested energy. When $p=1$, this regularizer shows no preference between EVs. For $p > 1$, EVs with higher $e_i$ will be prioritized (given more energy) over those with lower $e_i$ when it is infeasible to meet all energy demands. Note that this regularizer is 0 whenever the energy demands of all EVs are fully met, e.g. $\sum_{t \in \mathcal{T}} r_i(t) = e_i$. Thus, with sufficient weight on this component, \eqref{eq:Constraints.1c} will be tight whenever feasible. 
Likewise, if \eqref{eq:Constraints.1c} would have been tight without \eqref{eq:non-completion}, this regularizer has no effect. 
}

\subsection{Feasible set $\mathcal R_k$}
The feasible set $\mathcal R_k$ is defined by a set of equality and inequality constraints that can depend on $k$, but for notational simplicity, we drop the
subscript $k$.  These constraints then take the form:
\begin{subequations}
\begin{align}
& 0 \ \leq \ r_i(t) \ \leq \ \bar{r}_i(t) & t \leq d_i, i \in \mathcal{V}
\label{eq:Constraints.1a} \\
& r_i(t)  \ = \ 0  & t > d_i, i \in \mathcal{V}
\label{eq:Constraints.1b} \\
& \sum_{t\in\mathcal{T}} r_i(t) \ \leq \ e_i & i \in \mathcal{V}
\label{eq:Constraints.1c} \\
& \left| \sum_{i \in \mathcal{V}} A_{li} r_i(t) e^{j\phi_i} \right| \ \leq \ c_{lt}(t) & t \in \mathcal{T}, l \in \mathcal{L}
\label{eq:Constraints.1d}
\end{align}
\label{eq:Constraints.1}
\end{subequations}

\noindent Constraints (\ref{eq:Constraints.1a}) ensure that the charging rate in each period is non-negative (we do not consider V2G) and less than its upper bound defined by the EV's BMS and the maximum pilot supported by the EVSE.  This is a relaxation of the set of discrete rates allowed by the EVSE and is necessary to keep the scheduling problem convex. We discuss how to recover a feasible discrete solution in Section~\ref{sec:post_processing}. 
Constraints (\ref{eq:Constraints.1b}) ensure that an EV does not charge after its departure time. We use constraints (\ref{eq:Constraints.1c}) to limit the total energy delivered to EV $i$ to at most $e_i$. To ensure feasibility, we do not require equality (the zero vector is always a feasible solution). This ensures that \textbf{OPT} always returns a feasible schedule, which is important in practice. We can then craft the objective function to ensure this constraint is tight whenever possible%
\iflong{, see Section~\ref{sec:non-completion}}.

\subsection{Quantization of pilot signal} \label{sec:post_processing}
The pilot signal constraints imposed by EVSEs described in Section~\ref{sec:evse_limits} are discrete and intractable in general for large problems. Because of this, we do not include \eqref{eq:flapping-rate-const} in the definition of $\mathcal R_k$, instead relaxing it to \eqref{eq:Constraints.1a}. However, to account for our non-zero rate constraint, we add an additional constraint 
\begin{equation*}
    r_i(0) \geq \min \left( \rho_i(0) \setminus \{0\} \right)
\end{equation*}

\noindent to \eqref{eq:Constraints.1}.\footnote{This constraint implicitly assumes that it is feasible to deliver a minimum charging rate to each EV, thus charging infrastructure should be designed with this constraint in mind if the operators want to ensure a minimum charging rate to each EV.} We denote the output of this optimization $r^* := (r^*_i(t), \forall i\in\mathcal V \ \forall t\in\mathcal T)$. For simplicity, we assume that the maximum $P$ between scheduler calls (see Algorithm \ref{alg:MPC})
is set to the length of one period, so that only the first charging rate in $r^*$ will be applied.

We then round $r^*_i(0)$ down to the nearest value in $\rho_i$:
\begin{equation*} 
\tilde{r}_i(0) \ \leftarrow \ \left\lfloor{r^*_i(0)}\right\rfloor_{\rho_i}
\end{equation*}
\noindent This rounding may leave unused capacity which can be reclaimed. To reclaim this capacity, we first sort EVs in descending order by the difference between their originally allocated charging rate, $r^*_i(0)$, and the rate after rounding, $\tilde{r}_i(0)$. We then iterate over this queue and increment each EV's charging rate to the next highest value in $\rho_i(t)$, if the resulting current vector $\tilde{r}(0) \in \mathcal{R}_k$ and $\sum_i \tilde{r}_i(0) \leq \sum_i r^*_i(0)$. We continue to loop over this queue until we cannot increment any EV's allocated rate.

\iflong{
\subsection{Battery tail capacity reclamation} \label{sec:battery-tail-reclamation}
As discussed in Section~\ref{sec:bms_model}, an EV's battery management system will sometimes limit the power draw of the battery as it approaches 100\% state-of-charge.
When this happens, the difference between the pilot signal and the vehicle's actual charging rate is wasted capacity. To reclaim this capacity, we use a simple algorithm which we call \emph{rampdown}. Let $r^k_i(0)$ be the pilot signals sent to EV $i$ at time $k$, $m_i(k)$ be its measured charging current, and $\bar{r}_i^k(0)$ be the upper bound on its charging rate. We define two thresholds, $\theta_d$ and $\theta_u$. If $r_i^k(0) - m_i(k) > \theta_d$, we can reclaim some capacity by setting the upper limit on pilot signal of EV $i$ for the next period to be $m_i(k) + \sigma$, where $\sigma$ is typically around 1 A. In order to account for the possibility of the EV's BMS only limiting current temporarily, if $\bar{r}_i^k(0) - m_i(k) < \theta_u$, we increment the pilot signal upper bound by $\sigma$ (clipping at the EV's BMS limit or the EVSE's pilot limit). With this scheme, we can quickly reclaim capacity during the tail region, while still allowing EVs to throttle back up if this reclamation was premature. Note that in our current implementation, the upper bound on the pilot signal, $\bar r^k_i(t)$ is the same for all $t$ within the same sub-problem $k$. 
In more advanced rampdown schemes, this bound could depend on $t$ or the decision variables $r(t)$.
}

\section{Applications} \label{sec:applications}
We now turn our attention to applications of the Adaptive Charging Network. We first examine the real-world operational data we have collected from the system. We then use this data to evaluate (through simulations) how the Adaptive Scheduling Algorithm proposed in Section~\ref{sec:scheduling-framework} handles the practical challenges described in Section~\ref{sec:key-insights}. To do this, we consider two practical objectives, charging users quickly in highly constrained systems, and maximizing operating profits. 
\iflong{Due to limited space, we cannot address all possible use-cases of ACN and the ASA framework. For additional information about dynamic pricing and cost minimization using ACN see \cite{lee_pricing_2020}.}

\subsection{Data Collected}\label{sec:data_collected}
The ACN charging data includes actual arrival and departure times of EVs, estimated departure times and energy demands provided by drivers through the mobile app, and measurements of the actual energy delivered in each session. ACNs also record time series of the control signals passed to each EV and the EV's actual charging rate at 4-second resolution. This data is freely available, see \cite{lee_ACN-Data_2019}, and can be used to evaluate new scheduling algorithms using ACN-Sim. Important workload features include the arrival and departure distribution of EVs, shown in Fig.~\ref{fig:arrival_departure_dist}. 
\iflong{Other statistics are summarized in Table~\ref{tab:summary_stats}.}
This data was collected from May 1, 2018 - October 1, 2018 during which charging was free and only the 54 EVSEs connected to transformer $t_1$ were active. EVSEs on $t_2$ were added in early 2019, while the second and third garages in the ACN were added in late 2019. 
\iflong{During this period, the system served 10,415 charging sessions and delivered a total of 92.78 MWh of energy.}

\iflong{%
From Table \ref{tab:summary_stats}, we observe a significant difference in system usage between weekdays and weekends. The total energy delivered is much higher on weekdays, but this energy is divided over far more charging sessions leading to a lower per session energy delivery. Also, charging sessions on weekends tend to be shorter than those during the week. This means that our system must be able to handle large numbers of flexible sessions on weekdays and smaller numbers of relatively inflexible sessions on weekends. This behavior precludes simple solutions such as installing large numbers of level-1 chargers, which would be too slow on weekends and for low laxity weekday sessions, or small numbers of level-2 chargers, which would be insufficient for the number of concurrent sessions on weekdays.}

\begin{figure}[t!]
\includegraphics[width =\columnwidth]{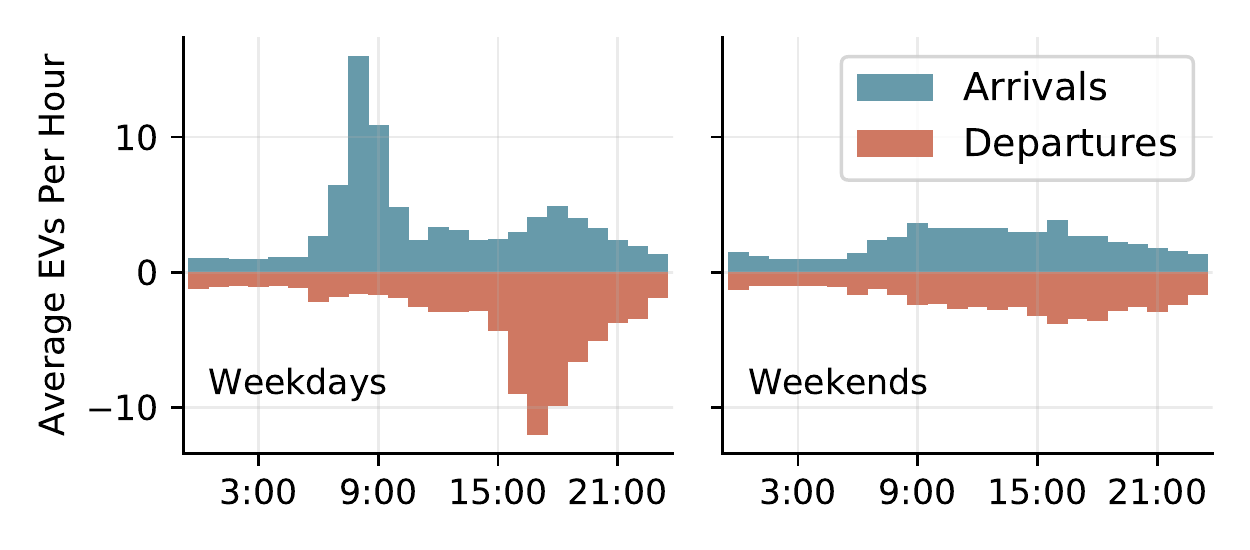}
\centering
\vspace{-0.2in}
\caption{Average arrivals and departures per hour for the period May 1, 2018 - October 1, 2018 (54 EVSEs). On weekdays we see a peak in arrivals between 7:00 - 10:00 followed by a peak in departures between 16:00 - 19:00. The Caltech ACN also has a much smaller peak in arrivals beginning around 18:00, which is made up of community members who use the site in the evening including some patrons of the nearby campus gym. Weekends, however have a more uniform distribution of arrivals and departures.}
\label{fig:arrival_departure_dist}
\end{figure}

\iflong{
\begin{table}[t!]
\centering
\captionsetup{justification=raggedright}
\caption{\hspace{15pt} Average Statistics for EV Charging Test Cases Per Day\newline
May 1, 2018 - Oct. 1, 2018 
}\label{tab:summary_stats}
\resizebox{\columnwidth}{!}{
\begin{tabular}{lccccc}
\toprule
  & \begin{tabular}[c]{@{}c@{}}Mean\\Daily\\Sessions \end{tabular} &
  \begin{tabular}[c]{@{}c@{}}Mean\\Session\\Duration\\(hours)\end{tabular} & \begin{tabular}[c]{@{}c@{}}Mean\\Session\\Energy\\(kWh)\end{tabular} & \begin{tabular}[c]{@{}c@{}}Mean\\Daily\\Energy\\(kWh)\end{tabular} &
  \begin{tabular}[c]{@{}c@{}}Max\\Concurrent\\Sessions\end{tabular}\\ \midrule
Sun & 41.32 & 3.94 & 10.05 & 415.06  & 18\\
Mon & 71.00 & 6.14 & 9.54 & 677.13  & 42\\
Tues & 76.73 & 6.24 & 8.94 & 685.79  & 47\\
Wed & 75.45 & 6.22 & 8.75 & 660.16  & 44\\
Thurs & 78.50 & 5.96 & 8.47 & 665.21 & 42\\
Fri & 77.18 & 6.71 &  9.04 & 697.41  & 43\\
Sat & 43.32 & 5.01 & 10.15 & 439.59  & 18\\ 
\midrule
\end{tabular}
}
\end{table}
}

\subsection{Practical Scenarios}
\begin{table}[t]
\caption{Modeling Assumptions by Scenario}
\label{tab:scenarios}
\centering
\begin{tabular}{@{}lccccc@{}}
\toprule
 & I & II & III & IV & V \\ \midrule
Perfect Information? & \cmark & \xmark & \xmark & \xmark & \xmark \\
Continuous EVSE? & \cmark & \cmark & \xmark & \cmark & \xmark \\
Ideal Battery? & \cmark & \cmark & \cmark & \xmark & \xmark \\ \midrule
\end{tabular}%
\end{table}

To better understanding the effect of practical limitations such as limited information, non-ideal batteries, and pilot signal quantization, we consider each operator objective in the context of the five scenarios in Table~\ref{tab:scenarios}. Here perfect information refers to having access to the arrival time, duration, and energy demand of all EVs in advance, allowing for offline optimization. Continuous EVSEs allow for continuous pilot control between 0 and the EVSE's upper bound, while quantized EVSEs only allow a discrete set of values and must keep the charging rate at or above 6 A until the EV is finished charging. Finally, ideal batteries are assumed to follow the pilot signal exactly. In contrast, non-ideal batteries follow the linear two-stage model described in \cite{lee_acnsim_2019}, where the initial state of charge and battery capacity are fit to maximize tail behavior, and the tail begins at 80\% state-of-charge.

For our simulations we use ACN-Sim, which includes realistic models for each of the scenarios above. In each case, we consider the three-phase infrastructure of the Caltech ACN. We set the length of each time slot to 5 minutes, the maximum time between scheduler calls to 5 minutes, and consider a maximum optimization horizon of 12 hours.  

\subsection{Energy delivery with constrained infrastructure}
We first consider the objective of maximizing total energy delivered when infrastructure is oversubscribed. This is a common use case when electricity prices are static or when user satisfaction is the primary concern. To optimize for this operator objective, we use the Adaptive Scheduling Algorithm (ASA) (Alg. \ref{alg:MPC}) with utility function

\begin{equation*}
    U^\mathrm{QC}(r) := u^{QC}(r) + 10^{-12}u^{ES}(r)
\end{equation*}

Here $U^{QC}$ encourages the system to deliver energy as quickly as possible, which helps free capacity for future arrivals. We include the regularizer $u^{ES}(r)$ to promote equal sharing between similar EVs and force a unique solution. We refer to this algorithm as ASA-QC.

To control congestion in the system, we vary the capacity of transformer $t_1$ between 20 and 150 kW. For reference, the actual transformer in our system is 150 kW, and a conventional system of this size would require 362 kW of capacity. We then measure the percent of the total energy demand met using ASA-QC as well as three baseline scheduling algorithms;  least laxity first (LLF), earliest deadline first (EDF), and round-robin (RR), as implemented in ACN-Sim.

Results from this experiment are shown in Fig.~\ref{fig:energy_delivered_const_infra}, from which we observe the following trends. 
\begin{enumerate}
    \item In scenario II, ASA-QC performs near optimally (within 0.4\%), and significantly outperforms the baselines (by as much as 14.1\% compared to EDF with 30 kW capacity).
    
    \item In almost all cases, ASA-QC performs better than baselines, especially so in highly congested settings.
    \footnote{For scenarios III and V and transformer capacities less than 68 kW, it may sometimes be infeasible to allocate a minimum of 6 A to each active EV. When this is the case, we allocate 6 A to as many EVs as possible then allocate 0 A to the rest.
    \iflong{This allocation is done by first sorting EVs (by laxity for LLF and ASA-QC, deadline for EDF, and arrival time for RR) then allocating 6 A to each EV until the infrastructure constraints are binding.}}
    
    \item Non-ideal EVSEs (scenarios III and V) have a large negative effect on ASA-QC, which we attribute to rounding of the optimal pilots and restriction of the feasible set.
    
    \item Surprisingly, non-ideal EVSEs increase the performance of LLF and EDF for transformer capacities $<$60 kW. This may be because the minimum current constraint leads to better phase balancing.
    
    \item Non-ideal batteries (scenarios IV and V) have relatively small effect on the performance of ASA-QC compared to baselines, indicating the robustness of the algorithm.
\end{enumerate}
\iflong{%
To understand why ASA-QC performs so much better than the baselines, especially in scenario II, we must consider what information each algorithm uses. RR uses no information aside from which EVs are currently present, and as such, performs the worst. Likewise, EDF uses only information about departure time, while LLF also makes use of the EVs energy demand. Only ASA-QC actively optimizes over infrastructure constraints, allowing it to better balance phases (increasing throughput) and prioritize EVs, including current and anticipated congestion. A key feature of the ASA framework is its ability to account for all available information cleanly.\footnote{When even more information is available, i.e., a model of the vehicle's battery or predictions of future EV arrivals, this information can also be accounted for in the constraint set $\mathcal{R}$ and objective $U(r)$. However, these formulations are outside the scope of this paper.}\\
}

\begin{figure}[t!]
\includegraphics[width=0.95\columnwidth]{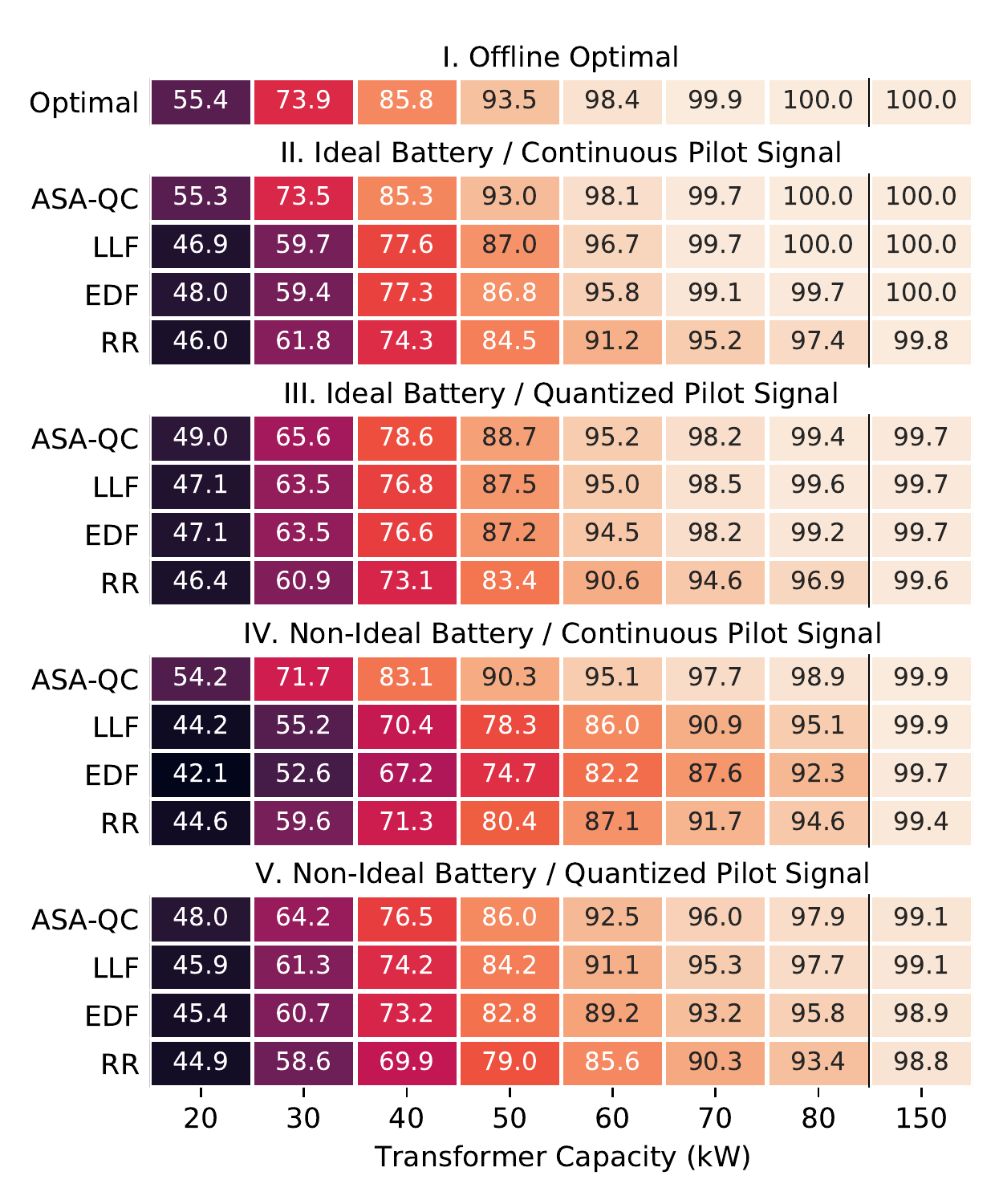}
\centering
\vspace{-0.2in}
\caption{Percentage of driver's energy demands that can be met at varying capacities for transformer $t_1$ for Sept. 2018. Here demand met is defined as the ratio of total energy delivered to total energy requested.
}
\label{fig:energy_delivered_const_infra}
\end{figure}

\subsection{Profit maximization with TOU tariffs and demand charge}\label{sec:cost_min}
Next, we consider the case where a site host would like to minimize their operating costs. Within this case, we will consider the Southern California Edison TOU EV-4 tariff schedule for separately metered EV charging systems between 20-500~kW, shown in Table~\ref{tab:tou_rates} \cite{choi_general_2017}. In each case, we assume that the charging system operator has a fixed revenue of \$0.30/kWh and only delivers energy when their marginal cost is less than this revenue. 
\begin{table}[t!]
\caption{SCE EV TOU-4 Rate Schedule for EV Charging (Summer)}
\label{tab:tou_rates}
\centering
\begin{tabular}{lccc}
\toprule
Name & Time Range & Weekday & Weekend\\ 
\midrule
Off-Peak & 23:00 -  8:00 &   \$0.056 / kWh   &   \$0.056 / kWh\\
\arrayrulecolor{black!30}\midrule
Mid-Peak & \begin{tabular}[c]{@{}l@{}} \ 8:00 - 12:00\\18:00 - 23:00\end{tabular}
&   \$0.092 / kWh   &   \$0.056 / kWh\\ 
\arrayrulecolor{black!30}\midrule
Peak &  12:00 - 18:00    &   \$0.267 / kWh   &   \$0.056 / kWh\\              
\arrayrulecolor{black!55}\midrule
Demand Charge  &  \multicolumn{3}{c}{\$15.51 / kW / month} \\ 
\arrayrulecolor{black}\bottomrule
\end{tabular}
\end{table}
In order to maximize profit, we use the objective:
\begin{align*}
    &U^\mathrm{PM} := u^{EC} + u^{DC} + 10^{-4}u^{QC} + 10^{ -12}u^{ES}
\end{align*}
\noindent We denote the ASA algorithm with this objective ASA-PM.
\ifelselong{%
The revenue term $\pi$ in $u^{EC}$ can have several interpretations. In the most straightforward case, $\pi$ is simply the price paid by users. However, $\pi$ can also include subsidies by employers, governments, automakers, or carbon credits through programs like the California Low-Carbon Fuel Standard (LCFS). For example, LCFS credits for EV charging have averaged between \$0.13 - \$0.16 / kWh in 2018-2019. In these cases, some energy demands might not be met if the marginal price of that energy exceeds $\pi$. This is especially important when demand charge is considered since the marginal cost can be extremely high if it causes a spike above the previous monthly peak. Alternatively, $\pi$ can be set to a very high value (greater than the maximum marginal cost of energy) and act as a non-completion penalty. When this is the case, the algorithm will attempt to minimize costs while meeting all energy demands (when it is feasible to do so).%
}
{%
The revenue term $\pi$ in $u^{EC}$ can be interpreted as either the price paid by the user, a subsidy paid by the operator, or as a price on unmet demand.
}

In $u^{DC}$, $\hat{P}$ and $q'$ are tunable parameters. The demand charge proxy $\hat{P}$ controls the trade-off between energy costs and demand charges in the online problem. In this case, we use the heuristic proposed in \cite{lee_pricing_2020}, $\hat{P} = P/(D_p - d)$, where $D_p$ is the number of days in the billing period, and $d$ is the index of the current day. We will consider one version of the algorithm without a peak hint, e.g. $q'=0$, and one where the peak hint is 75\% of the optimal peak calculated using data from the previous month. 
\iflong{This percentage is chosen based on maximum historic month-to-month variability in the optimal peak (+11\%/-16\%).}

We fix the transformer capacity to 150~kW and consider the previous baselines along with uncontrolled charging
\iflong{, which is the most common type of charging system today}.
Results of the experiment are shown in Fig.~\ref{fig:profit_max}, from which we observe:
\begin{enumerate}
    \item Profits from both ASA-PM and ASA-PM w/ Hint, are within 3.6\% and 1.9\% of the optimal respectively, and far exceed the profits of all baseline algorithms.
    
    \item Uncontrolled, LLF and RR result in \emph{lower} energy costs, but incur \emph{very high} demand charges. These algorithms are not price aware. Instead low energy costs are a result of drivers arriving during off-peak and mid-peak times.%
    \iflong{In particular, uncontrolled charging, which does not consider an infrastructure limit, leads to \emph{extremely high} demand charges. On the other hand, both ASA-PM algorithms (and the offline optimal) trade-off higher energy costs for much lower peaks resulting in lower overall costs.}
    
    \item Providing a peak hint to ASA-PM increases revenue by allowing more energy demands to be met. In this case, 97.8\% vs. 95.6\% without peak hints. Accurate hints allow the algorithm to utilize higher capacity earlier in the billing period, increasing throughput without increasing cost.
    \iflong{Even with the peak hint, ASA-PM does not meet 100\% of demands even though the offline optimal does. Since ASA-PM does not have knowledge of future arrivals, it must act conservatively in increasing the peak over time. It is, however, important that hints not be too large, as the algorithm can increase the peak as needed, but once a high peak is set, the demand charge cannot be lowered.}
    
    \item While EVSE quantization and non-ideal batteries each reduce the operator's profit, even in scenario V, MPC w/ Hint still produces 90\% of the optimal profit.
    
    \iflong{
    \item Interestingly, revenue increases in scenarios with quantization (III and V). It can be hard to reason about exactly why this occurs, though it appears that the post-processing step leads to initial conditions for the next solve of \textbf{OPT} to produce a higher revenue, higher cost solution. 
    }
    
    \item Because we use real tariffs structures, real workloads, and realistic assumptions (scenario V), we can conclude with reasonable certainty that a charging system operator could expect to net approximately \$2,600~/~month using an ACN like system, compared to just \$763~/~month in a conventional, uncontrolled system.
\end{enumerate}

\begin{figure}[t!]
\includegraphics[width = .85\columnwidth]{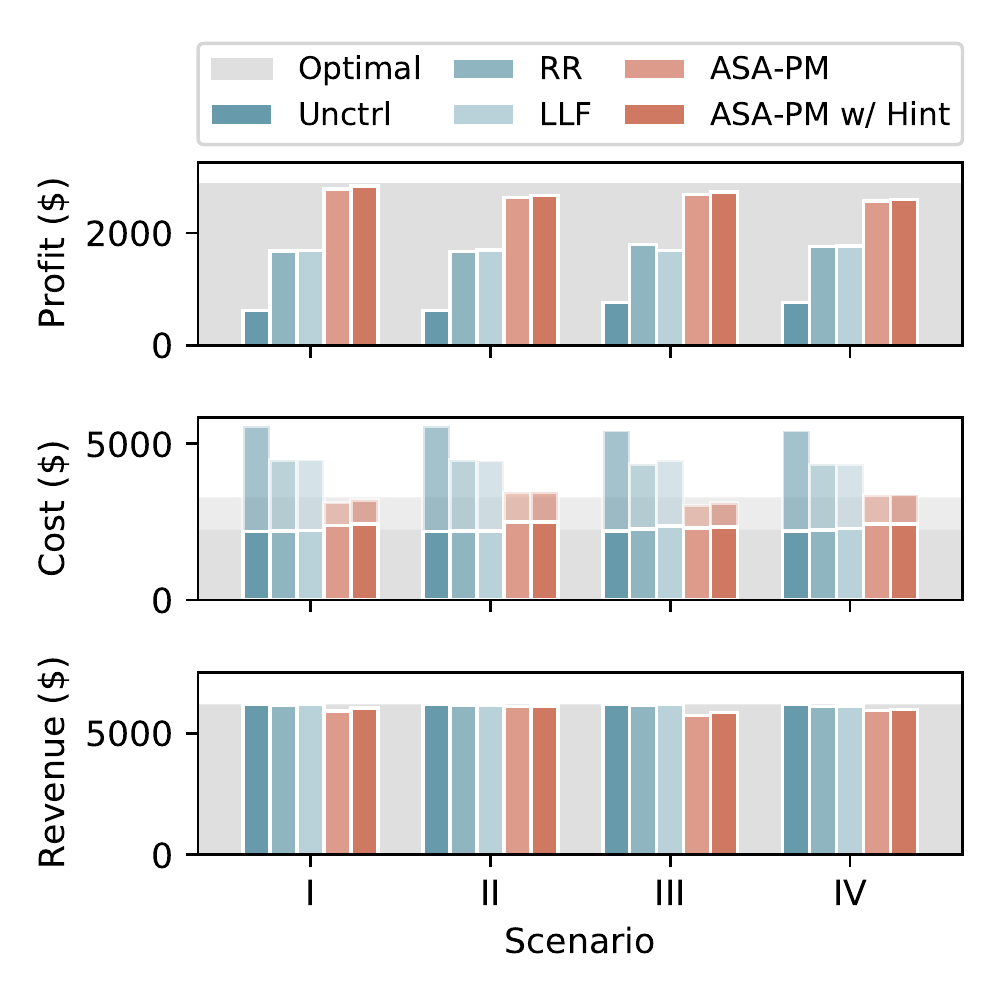}
\centering
\vspace{-0.25in}
\caption{Operator profit, costs, and revenue for various scheduling approaches when using SCE's EV TOU-4 tariff, $\pi$ = \$0.30, 150 kW transformer capacity, and data from Sept. 2018. In the middle panel we break out energy costs (darker, lower bar) from demand charge (lighter, upper bar). In each case, the offline optimal in the ideal setting is shown as a grey background. 
}
\label{fig:profit_max}
\end{figure}
\section{Conclusions} \label{sec-conclusions}
In this paper, we describe the Adaptive Charging Network (ACN), a framework for large-scale, managed electric vehicle charging facilities. The ACN and its scheduling algorithm (ASA) have been proven at scale through deployments around the United States, including the first ACN installed on the Caltech campus in 2016.

Through building the ACN, we have identified practical challenges, including unbalanced three-phase infrastructure, quantization of pilot signals, and non-ideal battery behavior, which require us to rethink classical scheduling approaches. To meet these challenges we propose ASA, a flexible model predictive control based algorithm along with pre- and post-processing heuristics, which can be easily configed to meet different operator objectives. 
\iflong{We propose a collection of such objectives, including regularizers to promote desirable properties in the final schedule.}

\ifelselong{%
Through case studies, we consider the objective of delivering as much energy as possible in constrained infrastructure and maximizing profit,  subject to time-of-use tariffs and demand charges. Using real workload data collected from the Caltech ACN and accurate models of ACN infrastructure, we demonstrate that ASA offers significant improvements in terms of energy delivered with constrained infrastructure when continuous pilots are allowed and performs comparably to baselines when pilots are restricted to a discrete set of values. We also note that by changing the objective function, we can easily modify ASA to maximize operator profit. Using real data from Sept. 2018, we achieve profits of \$2,835 (98.1\% of offline optimal) in an idealized setting, and \$2,600 (90\% of offline optimal) when considering non-ideal batteries and EVSEs. Compared to uncontrolled charging systems, our simulations show that an ACN like system can increase an EV charging system operator's profit by 3.4 times (see Section \ref{sec:cost_min})
}{%
Using real data from the ACN, we demonstrate that the ACN with ASA can reduce the infrastructure capacity necessary to meet charging demands and maximize profit subject to time-of-use tariffs and demand charges. In particular we find that ASA is able to consistently deliver more energy in highly constrained systems than baseline algorithms, and can increase an EV charging system operator's profit by 3.4 times over uncontrolled charging systems (see Section \ref{sec:cost_min}).
}

\iflong{
Beyond its operational role of charging hundreds of EVs each week, the Caltech ACN and similar sites at research institutions like JPL, NREL, SLAC, and UC San Deigo, provide a valuable platform for research in managed EV charging. To facilitate this new frontier of research, ACN Research Portal provides open-access data from the Caltech and JPL ACNs and an open-source simulation environment based on the ACN architecture. In addition, a new project called ACN-Live will allow researchers from anywhere in the world to field test algorithms on the Caltech ACN.
}
\iflong{\section{Acknowledgment}
Large-scale infrastructure projects like the ACN require herculean effort and coordination between researchers, administrators, private companies, and funding agencies. In addition to the authors, the ACN project would not have been possible without the efforts of John Onderdonk from Caltech Facilities, Stephanie Yankchinski and Neil Fromer from the Resnick Sustainability Institute, Jennifer Shockroo and Fred Farina from Caltech Office of Technology Transfer, and Roger Klemm from the JPL Green Club. Technology development was aided by data provided by Roff Schreiber of Google and generations of Caltech and international students since 2016, especially the work of Karl Fredrik Erliksson of Lund University. We would also like to thank PowerFlex Systems and EDF Renewables North America for their continued support, development, and commercialization of the ACN project.}{}

\bibliographystyle{ieeetr}
\scriptsize {
\bibliography{library,lib_extra}
}
\end{document}